\documentclass[aps,floatfix]{revtex4}

\usepackage{hyperref,hypcap,ae,tikz}

\usepackage{graphicx}% Include figure files
\usepackage{bm}% bold math
\usepackage{amsmath}
\usepackage[T1]{fontenc}
\usepackage{xcolor}
\usepackage{lmodern}
\usepackage[caption=false]{subfig}

\hypersetup{
	colorlinks=true,
	citecolor=blue,
    linkcolor=magenta,
	}

\begin{document}

\title{{\bf On the gravitational entropy of accelerating black holes }}

\author{\bf{Sarbari Guha and Samarjit Chakraborty}}

\affiliation{\bf Department of Physics, St.Xavier's College (Autonomous), Kolkata 700016, India}

\maketitle
\section*{Abstract}
In this paper we have examined the validity of a proposed definition of gravitational entropy in the context of accelerating black hole solutions of the Einstein field equations, which represent the realistic black hole solutions. We have adopted a phenomenological approach proposed in Rudjord et al [20] and expanded by Romero et al [21], in which the Weyl curvature hypothesis is tested against the expressions for the gravitational entropy. Considering the $C$-metric for the accelerating black holes, we have evaluated the gravitational entropy and the corresponding entropy density for four different types of black holes, namely, non-rotating black hole, non-rotating charged black hole, rotating black hole and rotating charged black hole. We end up by discussing the merits of such an analysis and the possible reason of failure in the particular case of rotating charged black hole and comment on the possible resolution of the problem.

\bigskip

KEYWORDS: Gravitational entropy, Accelerating Black holes.

\section{Introduction}
The $C$-metric was independently discovered by Levi-Civita \cite{Levi} and Weyl \cite{Weyl} in 1917. Ehlers and Kundt \cite{EK} while working on the classification of the degenerated static vacuum fields, constructed a table in which this metric was placed in the slot ``$C$'', leading to the name `$C$-metric'. Kinnersley and Walker \cite{KW} pointed out that this metric is an exact solution of Einstein's equations which describes the combined electromagnetic and gravitational field of a uniformly accelerating object having mass $m$ and charge $e$, and is an example of ``almost everything''. It is for this reason that the $C$-metric is the focus of our attention in this paper.

Dray and Walker \cite{DW} showed that this spacetime represents the gravitational field of a pair of uniformly accelerating black holes. Letelier and Oliveira \cite{LO}, studied the static and stationary $C$-metric and sought its interpretation in details, in particular those cases charaterized by two event horizons, one for the black hole and another for the acceleration. For spacetimes with vanishing or positive cosmological constant, the $C$-metric represents two accelerated black holes in asymptotically flat or de Sitter (dS) spacetime, and for a negative $\Lambda$ term, depending on the magnitude of acceleration \cite{DL}, it may represent a single accelerated black hole or a pair of causally separated black holes which accelerate away from each other \cite{Krtous}. The acceleration $A$ is due to forces represented by conical singularities arising out of a strut between the two black holes or because of two semi-infinite strings connecting them to infinity \cite{Podolsky,GKP}.

The second law of thermodynamics is one of the most fundamental laws of physics. We know that for an ensemble of ideal gas molecules confined to a closed chamber, the gas spreads out to fill the entire space once the chamber is opened, thereby reaching a state of maximum entropy. However, in the case of the universe with its matter content modelled as a fluid (or gas), this is not exactly true. The universe was born from a very homogeneous state and later on, small density fluctuations appeared due to the effect of gravity, that ultimately led to the formation of structures in the universe. This evolution is contrary to our expectations from the thermodynamic point of view, since the ``gas'' condenses into clumps of matter, instead of spreading out. Moreover in the past, the universe was much hotter and at some point of time, matter and radiation were in thermal equilibrium, and the entropy was maximum. So, how can the entropy increase if it was maximum in the past? It appears that if the evolution of the universe is dominated solely by gravity, then we may encounter a violation of the second law of thermodynamics, if we are considering the contribution of the thermodynamic entropy only.

To resolve this problem and to provide a proper sequence to the occurrence of gravitational processes, Penrose \cite{Penrose1} proposed that we must assign an entropy function to the gravitational field itself. He suggested that the Weyl curvature tensor could be used as a measure of the gravitational entropy. The Weyl tensor $C_{\alpha\beta\gamma\delta}$ in $ n $ dimensions is expressed as \cite{Chandra}
\begin{equation}\label{decom}
C_{\alpha\beta\gamma\delta}= R_{\alpha\beta\gamma\delta} - \dfrac{1}{(n-2)}(g_{\alpha\gamma}R_{\beta\delta} + g_{\beta\delta}R_{\alpha\gamma} - g_{\beta\gamma}R_{\alpha\delta} - g_{\alpha\delta}R_{\beta\gamma}) + \dfrac{1}{(n-1)(n-2)}R(g_{\alpha\gamma}g_{\beta\delta}-g_{\alpha\delta}g_{\beta\gamma}),
\end{equation}
where $R_{\alpha\beta\gamma\delta}$ is the covariant Riemann tensor, $R_{\alpha\beta}$ is the Ricci tensor and $R$ is the Ricciscalar.

According to Penrose, initially after the `big bang', when the universe started evolving, the Weyl tensor component was much smaller than the Ricci tensor component of the spacetime curvature. This hypothesis sounds credible because the Weyl tensor is independent of the local energy–momentum tensor. Moreover, the universe was in a nearly homogeneous state before structure formation began, and the FRW models successfully describe this homogeneous phase of the evolution. Further, the Weyl curvature is zero in the FRW models. However, the Weyl is large in the Schwarzschild spacetime. Thus we need a description of gravitational entropy, which should increase throughout the history of the universe on account of formation of more and more structures leading to the growth of inhomogeneity \cite{Penrose2,Bolejko}, and thus preserve the second law of thermodynamics. But there is still doubt regarding the definition of gravitational entropy in a way analogous to the thermodynamic entropy, which would be applicable to all gravitational systems \cite{CET}. The definition of gravitational entropy as the ratio of the Weyl curvature and the Ricci curvature faces problems with radiation \cite{Bonnor}. Once Senovilla showed that the Bel-Robinson tensor is suitable for constructing a measure of the ``energy'' of the gravitational field \cite{Senovilla}, several attempts were made to define the gravitational entropy based on the Bel-Robinson tensor and also in terms of the Riemann tensor and its covariant derivatives \cite{PL,PC}.

Many efforts has been made to explain the entropy of black holes using the quantized theories of gravity, such as the
string theory and loop quantum gravity. However, in this paper we will handle the problem from a phenomenological approach proposed in \cite{entropy1} and expanded in \cite{entropy2}, in which the Weyl curvature hypothesis is tested against the expressions for the entropy of cosmological models and black holes. They considered a measure of gravitational entropy in terms of a scalar derived from the contraction of the Weyl tensor and the Riemann tensor, and matched it with the Bekenstein-Hawking entropy \cite{SWH1,Bekenstein}. In our current work we will consider the accelerating black holes only, which represent more realistic black holes for several reasons. For instance, collision of galaxies is a rather common phenomenon occurring in the universe, and it inevitably leads to black hole mergers with the associated production of gravitational waves \cite{POK}. In such situations, we may imagine that the black holes at the centre of these galaxies are accelerating towards each other, although we can always think of any black hole as accelerating since no black hole is gravitationally isolated from the neighboring massive systems. Moreover, a static black hole may be considered as the limiting case of an accelerating black hole. Thus the study of accelerating black holes is very important. Here we will investigate whether the calculations for gravitational entropy proposed in \cite{entropy1} and \cite{entropy2} can be applied in this context. The organization of our paper is as follows: Sec. II deals with the definition of gravitational entropy and Sec. III enlists the metrics of accelerating black holes considered by us. Sec. IV provides the main analysis of our paper where we evaluate the gravitational entropy and the corresponding entropy density for these black holes. We discuss our results in Sec. V and present the conclusions in Sec. VI.

\section{Gravitational Entropy}
The entropy of a black hole can be described by the surface integral \cite{entropy1}
\begin{equation}
S_{\sigma}=k_{s}\int_{\sigma}\mathbf{\Psi}.\mathbf{d\sigma},
\end{equation}
where $ \sigma $ is the surface of the horizon of the black hole and the vector field $\mathbf{\Psi}$ is given by
\begin{equation}
\mathbf{\Psi}=P \mathbf{e_{r}},
\end{equation}
with $ \mathbf{e_{r}} $ as a unit radial vector. The scalar $ P $ is defined in terms of the Weyl scalar ($ W $) and
the Krestchmann scalar ($ K $) in the form
\begin{equation}\label{P_sq}
P^2=\dfrac{W}{K}=\dfrac{C_{abcd}C^{abcd}}{R_{abcd}R^{abcd}}.
\end{equation}
In order to find the gravitational entropy, we need to do our computations in a 3-space. Therefore, we consider the spatial metric which is defined as
\begin{equation}\label{sm}
h_{ij}=g_{ij}-\dfrac{g_{i0}g_{j0}}{g_{00}},
\end{equation}
where $ g_{\mu\nu} $ is the concerned 4-dimensional space-time metric and the Latin indices denote spatial components, $i, j = 1, 2, 3$. So the infinitesimal surface element is given by
\begin{equation}
d\sigma=\dfrac{\sqrt{h}}{\sqrt{h_{rr}}}d\theta d\phi.
\end{equation}
Using Gauss's divergence theorem, we can easily find out the entropy density \cite{entropy1} as
\begin{equation}
s=k_{s}|\mathbf{\nabla}.\mathbf{\Psi}|.
\end{equation}

\section{Accelerating Black holes}

\subsection{Non-rotating black hole}
The $C$-metric in spherical type coordinates is given by
\begin{equation}\label{cmetric}
ds^2=\dfrac{1}{(1-\alpha r cos\theta)^2}\left(-Qdt^2+\dfrac{dr^2}{Q}+\dfrac{r^2d\theta^2}{P}+Pr^2sin^2\theta d\phi^2\right),
\end{equation}
where $ P=(1-2\alpha m cos\theta)$, and $ Q=\left(1-\dfrac{2m}{r}\right)(1-\alpha^2r^2) $. This metric represents an accelerating massive black hole, which has two coordinate singularities, one is at $ r_{a}=\dfrac{1}{\alpha} $ and the other is at $ r_{h}=2m $. The $ r_{h}=2m $ singularity stands for the familiar \emph{event horizon}, but the $ r_{a}=\dfrac{1}{\alpha} $ singularity is the \emph{acceleration horizon} formed due to the acceleration of the black hole \cite{GP}. Here $m$ is the mass of the black hole and $ \alpha $ is the acceleration parameter. The important feature of this metric is that  $ \phi\in [0,2\pi C) $ unlike the full $ 2\pi $ range for stationary black holes because of the conical singularity arising due to acceleration. Here $ C=\dfrac{1}{(1+2\alpha m)} $ is the deficiency factor in the range of $ \phi $. If the acceleration of the black hole vanishes, i.e., $ \alpha=0 $, then the deficiency factor $ C $ becomes unity and $ \phi $ reduces to the conventional polar coordinate running from $ 0 $ to $ 2\pi $. All the accelerated black hole metrics discussed below also have this property.

\subsection{Non-rotating charged black hole}
We now consider the metric representing charged accelerating black holes. The charged $C$-metric in spherical type coordinate is \cite{GP}
\begin{equation}\label{cmetric1}
ds^2=\dfrac{1}{(1-\alpha r cos\theta)^2}\left(-Qdt^2+\dfrac{dr^2}{Q}+\dfrac{r^2d\theta^2}{P}+Pr^2sin^2\theta d\phi^2\right),
\end{equation}
where $ P=(1-2\alpha m cos\theta+\alpha^2e^2cos^2\theta)$, and $ Q=\left(1-\dfrac{2m}{r}+\dfrac{e^2}{r^2}\right)(1-\alpha^2r^2) $.
This is just the charged version of the previous metric with the parameter $ e $ representing the charge of the black hole. We can also think of it as an accelerated Reissner–Nordstrom (RN) black hole, because as $ \alpha\rightarrow 0 $, the metric reduces to the familiar RN metric. In the case of this metric also, we have $ r=\dfrac{1}{\alpha} $ as the acceleration horizon, and because of the introduction of charge, we have the outer and inner horizons at $ r_{\pm}=m\pm \sqrt{m^{2}-e^{2}}.$ Here the corresponding deficiency factor is given by $ C=\dfrac{1}{(1+2\alpha m+\alpha^2 e^2)} $.

\subsection{Rotating black hole}
The general line element for \textbf{an accelerating} rotating black hole is given by
\begin{equation}\label{htn}
ds^2=\dfrac{1}{\Omega^2}\left(-\dfrac{Q}{R}(dt-asin^2\theta d\phi)^2 + \dfrac{R}{Q}dr^2+\dfrac{R}{P}d\theta^2+\dfrac{P}{R}sin^2\theta[adt-(r^2+a^2)d\phi]^2\right),
\end{equation}
where $ \Omega=1-\alpha rcos\theta $, $ R=r^2+a^2cos^2\theta $, $ P=(1-2\alpha m cos\theta +\alpha^2a^2cos^2\theta) $, and $ Q=(a^2-2mr+r^2)(1-\alpha^2r^2) $. This metric represents the rotating version of the $ C$-metric, and contains three coordinate singularities, namely $ r_{\pm}=m\pm \sqrt{m^2-a^2} ,$ representing the outer and inner horizons, and $ r=\dfrac{1}{\alpha} $ representing the acceleration horizon \cite{GP} with the deficiency factor $C=\dfrac{1}{(1+2\alpha m+\alpha^2 a^2)}$ .

\subsection{Rotating charged black hole}
This is the charged version of the previous accelerating rotating metric. It may be regarded as the most general case among all the black holes considered by us, and is given by
\begin{equation}\label{ht}
ds^2=\dfrac{1}{\Omega^2}\left(-\dfrac{Q}{R}(dt-asin^2\theta d\phi)^2 + \dfrac{R}{Q}dr^2+\dfrac{R}{P}d\theta^2+\dfrac{P}{R}sin^2\theta[adt-(r^2+a^2)d\phi]^2\right),
\end{equation}
where $ \Omega=1-\alpha rcos\theta $, $ R=r^2+a^2cos^2\theta $, $ P=(1-2\alpha m cos\theta +\alpha^2(a^2+e^2)cos^2\theta) $, and $ Q=(a^2+e^2-2mr+r^2)(1-\alpha^2r^2) $. In this case the deficiency factor is given by $ C=\dfrac{1}{(1+2\alpha m+\alpha^2 (a^2+e^2))}. $ As in the previous case, the acceleration horizon is at $ r=\dfrac{1}{\alpha} $, however, the outer (or inner) horizons are located at $ r_{\pm}=m\pm \sqrt{m^2-a^2-e^2}. $

\section{Analysis}

\subsection{Non-rotating accelerating black hole}

We know that the Kretschmann scalar for a given spacetime geometry is defined by the relation
\begin{equation}
K=R_{abcd}R^{abcd},
\end{equation}
where $R_{abcd} $ is the covariant Riemann curvature tensor. For the $C$-metric (\ref{cmetric}), the Kretschmann scalar turns out to be
\begin{equation}\label{non-rot_Kscalar}
K_{c}=\dfrac{48m^2(\alpha r cos\theta-1)^6}{r^6}.
\end{equation}
The Weyl scalar is defined by
\begin{equation}
W=C_{abcd}C^{abcd},
\end{equation}
where the $C_{abcd} $ is the  Weyl curvature tensor. For the $C$-metric (non-rotating black hole) the Weyl scalar is evaluated as
\begin{equation}\label{non-rot_Wscalar}
W_{c}=\dfrac{48m^2(\alpha r cos\theta-1)^6}{r^6}.
\end{equation}
This result is expected since the Ricci tensor for this metric turns out to be zero. As the Riemann tensor can be decomposed into the Ricci and the Weyl parts according to equation (\ref{decom}), the vanishing Ricci component renders the Riemann and Weyl tensors identical as evident from equations (\ref{non-rot_Kscalar}) and (\ref{non-rot_Wscalar}).
The scalar function $P$ is defined by the relation (\ref{P_sq}) as
\begin{equation}
P^2=\dfrac{C_{abcd}C^{abcd}}{R_{abcd}R^{abcd}}.
\end{equation}
For this $C$-metric, we get $ P^2=1 $. Therefore we assume that $ P=+1 $ for our entropy calculations, since the entropy must be non-negative.

Now the \emph{spatial section} corresponding to this metric is
\begin{equation}
 h_{ij}=diag\left[\dfrac{1}{( 1 - \alpha r cos\theta )^2 (1-2m/r) (1-\alpha^2 r^2)},
\dfrac{r^2}{(( 1 - \alpha r cos\theta )^2 ( 1 - 2\alpha m cos\theta ))},
\dfrac{r^2 sin^2 \theta ( 1 - 2 \alpha m cos\theta )}{(1 - \alpha r cos\theta )^2}\right],
\end{equation}
with the determinant given by
\begin{equation}
h=\dfrac{sin^2(\theta)r^5}{(\alpha^2 r^2-1)(-r+2m)(\alpha r cos(\theta)-1)^6}.
\end{equation}
Therefore, the infinitesimal surface element has the form
\begin{equation}
d\sigma=\dfrac{\sqrt{h}}{\sqrt{h_{rr}}}d\theta d\phi=\dfrac{r^2 sin\theta}{(\alpha r cos\theta-1)^2} d\theta d\phi.
\end{equation}

\begin{figure}
\includegraphics[width=0.34\textwidth]{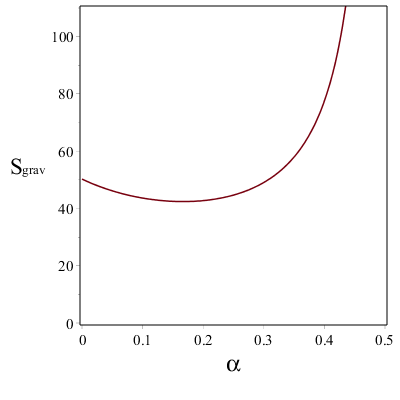}
\caption{Plot showing the variation of the total gravitational entropy for the accelerating non-rotating BH with respect to the acceleration parameter $ \alpha $, where we have taken $m=1 \: \textrm{and} \: k_{s}=1$.}\label{plot3}
\end{figure}

We are now in a position to calculate the magnitude of the gravitational entropy on the event horizon $H_{0}$ at the location $ r_{h}=2m $ for this metric, which is
\begin{equation}\label{s_grav_nonrot}
S_{grav}=k_{s}r_{h}^2\int_{\theta=0}^{\pi}\dfrac{sin\theta}{(\alpha r_{h} cos\theta-1)^2}d\theta \int_{\phi=0}^{2\pi C} d\phi=k_{s}\dfrac{4 \pi C r_{h}^2}{(1-r_{h}^2\alpha^2)}=k_{s}\dfrac{4 \pi r_{h}^2}{(1-r_{h}^2\alpha^2)(1+2\alpha m)}.
\end{equation}
From equation (\ref{s_grav_nonrot}) it is evident that the gravitational entropy is proportional to the area of the event horizon of the black hole, as in the case of the Bekenstein-Hawking entropy \cite{SWH1,Bekenstein}. Here $ C=\dfrac{1}{(1+2\alpha m)} $ is the deficiency factor in the limit of $ \phi $ as it runs from $ 0\rightarrow 2\pi C $ (as mentioned earlier). In FIG. \ref{plot3}, we have shown the variation of the total entropy on the horizon with the acceleration parameter $ \alpha $.

\begin{figure}[ht]
    \centering
    \subfloat[Subfigure 1 list of figures text][]
        {
        \includegraphics[width=0.4\textwidth]{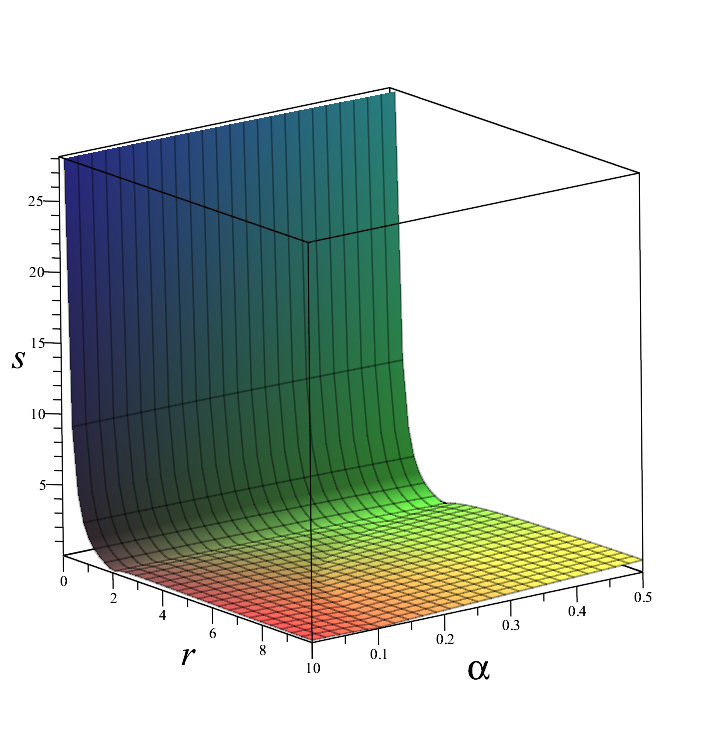}
        \label{fig:subfig1}
        }
    \hspace{0.5cm}
    \subfloat[Subfigure 2 list of figures text][]
        {
        \includegraphics[width=0.34\textwidth]{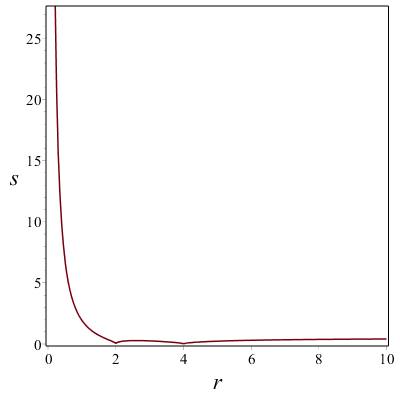}
        \label{fig:subfig2}
        }
    \caption{(a) Plot showing the variation of the gravitational entropy density for an accelerating non-rotating BH with respect to the acceleration parameter $ \alpha $ and the radial coordinate $ r $, for $ m=1$ and $ k_{s}=1 $. (b) Plot showing the variation of the gravitational entropy density  for the accelerating non-rotating BH with respect to the radial coordinate $ r $, where $ \alpha=0.25$, $ m=1$, and $ k_{s}=1 $.}
    \label{fig2}
\end{figure}

Similarly we can compute the entropy density as
\begin{equation}\label{s_nonrot}
s=k_{s}\frac{1}{\sqrt{h}}\frac{\partial}{\partial r}\left(\sqrt{h}\dfrac{P}{\sqrt{h_{rr}}} \right)=\dfrac{2k_{s}}{r}\sqrt{\left(1-\alpha^{2}r^{2}\right)\left(1-\frac{r_{h}}{r}\right)}.
\end{equation}
In the above equation (\ref{s_nonrot}), inserting $ \alpha=0 $, we get the entropy density for the Schwarzschild black hole. In FIG. \ref{fig2}, the dependence of the gravitational entropy density corresponding to this metric on other relevant parameters have been indicated. From equation (\ref{s_nonrot}) we can see that the zeroes of the gravitational entropy density function are located at the acceleration horizon $ r=\dfrac{1}{\alpha} $, and at the event horizon $ r=2m $, which is clearly evident from FIG. \ref{fig2}. Specifically, FIG. \ref{fig2}(a) shows that for $ \alpha=0 $, the acceleration horizon goes to infinity where the entropy density reduces to zero, and at the event horizon $r=2$, the entropy density becomes zero. Similarly for $ \alpha=0.5 $, the acceleration horizon and the event horizon coincide at $ r=2 $, where the entropy density becomes zero. FIG. \ref{fig2}(b) indicates that for $\alpha=0.25$, the acceleration horizon lies at $r=4$ and the event horizon is at $ r=2 $, the entropy density going to zero at both these places, and diverges at the singularity $ r=0 $ which is in agreement with equation (\ref{s_nonrot}).

\subsection{Non-rotating charged accelerating black hole}

The Kretschmann scalar for the non-rotating charged black hole given by the metric (\ref{cmetric1}) is evaluated to be
\begin{equation}
K=\dfrac{56\left(\alpha rcos\theta-1\right)^6 \left(cos^2\theta \alpha^2 e^4 r^2+\dfrac{10}{7} \left(e^2-\dfrac{6}{5}mr\right)re^2 \alpha cos\theta+e^4-\dfrac{12}{7} e^2mr+\dfrac{6}{7}m^2r^2 \right)}{r^8},
\end{equation}
and the corresponding Weyl scalar is
\begin{equation}
W=\dfrac{4}{3}\dfrac{\left(\alpha r cos\theta -1\right)^4 \left(5cos^2\theta \alpha^2 e^2 r^2-sin^2\theta\alpha^2 e^2 r^2+\alpha^2 e^2 r^2-6 m \alpha cos\theta r^2-6e^2+6mr\right)^2}{r^8}.
\end{equation}

Therefore the quantity $P$ is given by the expression
\begin{equation}
P^2=\dfrac{6(e^2\alpha cos\theta r+e^2-mr)^2}{(7 cos^2\theta\alpha^2e^4r^2+10\alpha r cos\theta e^4-12\alpha r^2 cos\theta e^2 m+7e^4-12e^2 mr+6 m^2 r^2)}.
\end{equation}

The spatial metric for this case is
\begin{eqnarray}
  h_{ij} &=& \dfrac{1}{(1-\alpha r cos\theta)^2} diag\left[\dfrac{1}{\left(1-\dfrac{2m}{r}+\dfrac{e^2}{r^2}\right)\left(-\alpha^2 r^2+1 \right)},
   \dfrac{r^2}{\beta}, \beta r^2 sin^2\theta \right],
\end{eqnarray}
where
\begin{equation}
\beta=\left(1-2\alpha m cos\theta +\alpha^2 e^2 cos^2\theta\right). \nonumber
\end{equation}
Consequently the determinant of the spatial $ h_{ij} $ metric is given by
\begin{equation}
h=-\dfrac{sin^2\theta r^6}{(\alpha^2 r^2-1)(e^2-2mr+r^2)(\alpha r cos\theta-1)^6},
\end{equation}
and the infinitesimal surface element is
\begin{equation}
d\sigma=\dfrac{\sqrt{h}}{\sqrt{h_{rr}}}d\theta d\phi=\dfrac{r^2 sin\theta}{(\alpha r cos\theta-1)^2} d\theta d\phi.
 \end{equation}

Next we calculate the gravitational entropy on the horizon $H_{0}$ at $ r_{h}=r_{\pm}=m \pm \sqrt{m^2-e^2} ,$ which turns out to be
\begin{equation}\label{s_grav_nonrot_chrg}
S_{grav}=k_{s}r_{h}^2\int_{\theta=0}^{\pi}\dfrac{P(r_{h},\theta) sin\theta}{(\alpha r_{h} cos\theta-1)^2}d\theta \int_{\phi=0}^{2\pi C} d\phi=\dfrac{k_{s}(4\pi r_{\pm}^2)}{(1+2\alpha m+\alpha^2 e^2)}\int_{\theta}\dfrac{P_{\pm}(\theta)sin\theta d\theta}{2(\alpha r_{\pm} cos\theta-1)^2},
\end{equation}
where the quantity $P_{\pm}$ corresponds to the value calculated for $r_{\pm}$.

\begin{figure}[ht]
    \centering
    \subfloat[Subfigure 1 list of figures text][]
        {
        \includegraphics[width=0.36\textwidth, height=0.3\textheight]{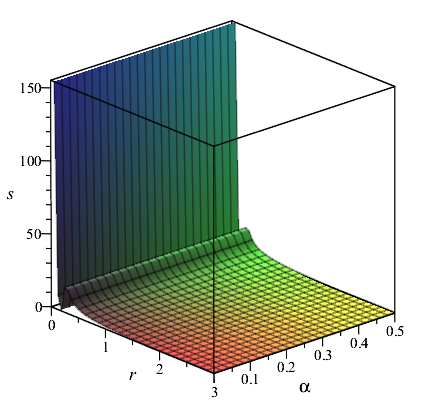}
        \label{fig:subfig3}
        }
        \hspace{1.0cm}
    \subfloat[Subfigure 2 list of figures text][]
        {
        \includegraphics[width=0.36\textwidth, height=0.32\textheight]{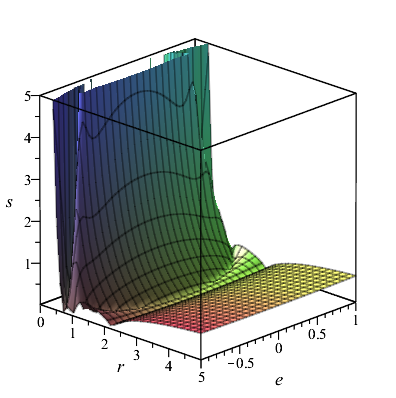}
        \label{fig:subfig4}
        }
    \caption{(a) Plot showing the variation of the gravitational entropy density for the accelerating non-rotating charged BH with respect to the acceleration parameter $ \alpha $ and the radial coordinate $ r $, where $ m=1, \; k_{s}=1, \; e=0.5, \; \textrm{and} \; \theta=\dfrac{\pi}{2} $. (b) Plot showing the variation of the gravitational entropy density for the accelerating non-rotating charged BH with respect to the radial coordinate $ r $ and the charge $ e $, where $ \alpha=0.45, \: m=1, \: k_{s}=1, \: \textrm{and} \: \theta=\dfrac{\pi}{2} $.}
    \label{fig3}
\end{figure}

\begin{figure}
\includegraphics[width=0.55\textwidth]{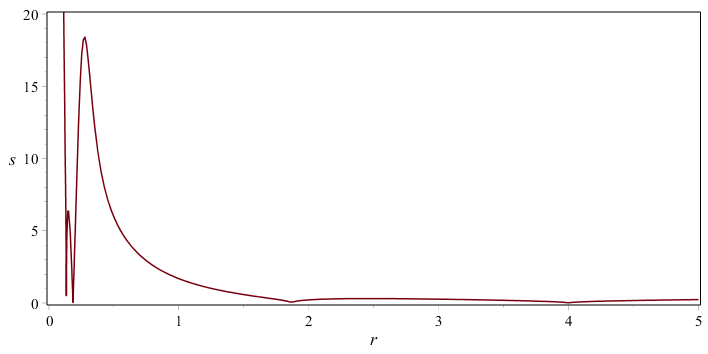}
\caption{Plot showing the variation of the gravitational entropy density for the accelerating non-rotating charged BH with respect to the  radial coordinate $ r $, where $ m=1, \; k_{s}=1, \; \alpha=0.25, \;  e=0.5, \; \textrm{and} \; \theta=\dfrac{\pi}{2} $.}\label{plot4a}
\end{figure}

From equation (\ref{s_grav_nonrot_chrg}) we find that the gravitational entropy is proportional to the area of the event horizon of the black hole, just as in the case of the Bekenstein-Hawking entropy. We can further check the validity of our result by setting $\alpha=0$ in (\ref{s_grav_nonrot_chrg}), to see whether it leads us to the desired expression for the entropy of the Reissner–Nordstrom (RN) black hole. This exercise yields the result
\begin{equation}
S_{grav}^{RN}=k_{s}(4\pi r_{\pm}^{2})\int_{\theta}P^{RN}_{\pm}(\theta)\dfrac{\sin\theta}{2} d\theta.
\end{equation}
We can easily see that $P^{RN}_{\pm}(\theta)=P_{\pm}(\alpha=0)=\dfrac{6e^{4}-12e^{2}mr+6m^{2}r^{2}}{7e^{4}-12e^{2}mr+6m^{2}r^{2}} , $ and
therefore the gravitational entropy for the RN black hole is
\begin{equation}
S_{grav}^{RN}=k_{s}(4\pi r_{\pm}^{2})\sqrt{\dfrac{6e^{4}-12e^{2}mr+6m^{2}r^{2}}{7e^{4}-12e^{2}mr+6m^{2}r^{2}}}\int_{\theta}\dfrac{\sin\theta}{2}d\theta=k_{s}(4\pi r_{\pm}^{2})\sqrt{\dfrac{6e^{4}-12e^{2}mr+6m^{2}r^{2}}{7e^{4}-12e^{2}mr+6m^{2}r^{2}}}.
\end{equation}
This result matches with the expression of gravitational entropy for the RN black hole derived in \cite{entropy2} by Romero et al.
The entropy density for the non-rotating charged black hole is obtained as
\begin{align} \label{s_nonrot_chrg}
\left.s\right. & = \frac{16\sqrt {6}k_{s}\sqrt { \left( -{\alpha}^{2}{r}^{2}+1 \right)
 \left( {e}^{2}-2mr+{r}^{2} \right) }}{{{r}^{2} \left( 7{e}^{4}{\alpha}^{2} \cos^{2}\theta{r}^{2} + 10 \left( {e}^{2}-\dfrac{6mr}{5} \right) r\alpha{e}^{2}\cos\theta + 7{e}^{4}-12{e}^{2}mr+6{m}^{2}{r}^{2} \right) ^{3/2}}}   \nonumber \\
& \times \left[ \cos^{3}\theta{\alpha}^{3}{e}^{6}{r}^{3} + {\frac {15{e}^{4}{\alpha}^{2} \cos^{2}\theta{r}^{2}}{8} \left( {e}^{2} - {\frac{13mr}{10}} \right) } + \dfrac{9 r\alpha{e}^{2} \cos\theta }{4} \left( {e}^{4}-{\frac {11{e}^{2}mr}{6}}+{m}^{2}{r}^{2} \right) \right. \nonumber \\
 & \qquad\qquad\qquad + \left. {\frac {7{e}^{6}}{8}} - \frac{3mr}{4} \left({ \frac {13{e}^{4}}{4}} - 3{e}^{2}mr + {m}^{2}{r}^{2} \right) \right] .
 \end{align}
If in this expression we substitute $ e=0 $, and consider the absolute value of this quantity, then we get back the expression (\ref{s_nonrot}) for the entropy density of the accelerating black hole. In FIG. \ref{fig3} we have shown the dependence of the gravitational entropy density of the non-rotating charged black hole on different parameters appearing in (\ref{s_nonrot_chrg}). From FIG. \ref{fig3}(a) we can determine the zeroes of the gravitational entropy function (\ref{s_nonrot_chrg}), e.g., for $ \alpha=0, \theta=\frac{\pi}{2} $, the acceleration horizon goes to infinity and by solving the entropy density function, we obtain the zeroes at $ r=0.13, 1.87 $, and also at $ r=0.18 $, where $ r=0.13, 1.87 $ are the horizons. Again from (\ref{s_nonrot_chrg}), using $ \alpha=0.45, \theta=\frac{\pi}{2} $, we find that the zeroes of the entropy density function are located at the acceleration horizon $ r=\frac{1}{\alpha}=2.22 $, and at the event horizon $ r=m\pm\sqrt{m^2 - e^2}=1 $, which is evident from FIG. \ref{fig3}(b). The additional zero can be found by solving for the roots of the second factor in (\ref{s_nonrot_chrg}) which gives us the only real root at $ r=0.74 $.  We have also analyzed the case for $ \alpha=0.25 $ (shown in FIG. \ref{plot4a}) from which we can identify the zeroes clearly, i.e., at the acceleration horizon $ r=\frac{1}{\alpha}=4 $, and at the horizons $ r=0.13, 1.87 $. Further, another zero arises from the second term of the entropy density function at $ r=0.19 $. The overall behavior is also as we expected, that is, the entropy density diverges near the $r=0$ singularity and it increases inside the horizon, encountering some zeroes in between.

\subsection{Rotating accelerating black hole}
Using the metric for the rotating accelerating black hole we have calculated the Ricci tensor, which turns out to be zero. Thus the Kretschmann scalar $K$ and the Weyl scalar $W$ are identical. Thus we have
\begin{eqnarray}
\nonumber
K= W = 48{m}^{2} \left( \alpha r\cos \left( \theta \right) -1 \right) ^{6} \left( \left({a}^{4}\alpha+{a}^{3} \right) \cos^{3}\theta
+ 3{a}^{2}r \left( a \alpha-1 \right) \cos^{2}\theta - 3a {r}^{2} \left( a\alpha+1 \right) \cos \theta -{r}^{3}  \left( a\alpha-1 \right)  \right)  \\
\times \dfrac{ \left(  \left( {a}^{4}\alpha - {a}^{3} \right) \cos^{3}\theta - 3{a}^{2}r \left( a \alpha+1 \right)  \cos^{2}\theta
- 3a {r}^{2}\left(a\alpha-1 \right) \cos \theta + {r}^{3}  \left( a\alpha+1 \right)  \right) } { \left( {r}^{2} + {a}^{2} \cos^{2}\theta \right) ^{6} }.
\end{eqnarray}
Therefore $ P^{2}=\dfrac{W}{K}=1 $, i.e. $ P=+1 $. Hence the total gravitational entropy in this case is given by
\begin{equation}\label{s_grav_rot}
S_{grav}=k_{s}\int_{\sigma}\mathbf{\Psi}.\mathbf{d\sigma}=k_{s}\int_{\sigma}d\sigma=k_{s}\int_{\theta=0}^{\pi}\int_{\phi=0}^{2\pi C}\sqrt{g_{\theta\theta}g_{\phi\phi}}d\theta d\phi.
\end{equation}
The entropy evaluated at $r_{\pm}$ is obtained as
\begin{equation}\label{s_grav_pm}
S_{grav_{\pm}}=k_{s}\dfrac{4\pi C(r^{2}_{\pm}+a^{2})}{(1-\alpha^{2}r_{\pm}^{2})}=k_{s}\dfrac{4\pi (r^{2}_{\pm}+a^{2})}{(1-\alpha^{2}r_{\pm}^{2})(1+2\alpha m+\alpha^2a^2)}.
\end{equation}
If we substitute $ a=0 $ in (\ref{s_grav_pm}), then we get back the expression (\ref{s_grav_nonrot}) for the entropy of the non-rotating accelerating black holes. We see that as the acceleration parameter vanishes, i.e., $ \alpha\rightarrow 0 $, the equation (\ref{s_grav_pm}) reduces to the expression of gravitational entropy for Kerr black holes  derived in \cite{entropy2}. However for this axisymmetric metric, it is not possible to evaluate the spatial metric using equation (\ref{sm}) because the object is rotating, and so there is a nonzero contribution from the component of $ g_{t\phi} $, which changes the spatial positions of events in course of time. Therefore the entropy density is calculated by using the full four-dimensional metric determinant $ g $ in the expression involving the covariant derivative \cite{entropy2}, and we get
\begin{equation}\label{enden1}
s=k_{s}|\mathbf{\nabla}.\mathbf{\Psi}|=\dfrac{k_{s}}{\sqrt{-g}}\left(\dfrac{\partial}{\partial r}\sqrt{-g}P\right)=2k_{s}\dfrac{(2\cos^{3}\theta a^{2}\alpha+\cos\theta \alpha r^{2}+r)}{(1-\alpha r \cos\theta)(r^{2}+a^{2}cos^{2}\theta)},
\end{equation}
where $ g=-\sin^{2}\theta \dfrac{(a^{2}\cos^{2}\theta+r^{2})^{2}}{(\alpha r \cos\theta -1)^{8}} $.

From equation (\ref{enden1}) we see that the entropy density diverges at the ring singularity and at $ r=\dfrac{1}{\alpha\cos\theta}, $ which is the conformal infinity in this spherical type coordinate system, as is evident from the metric (\ref{htn}). This can also be further verified from the expressions of the Kretschmann scalar and the Weyl scalar in this case, since they vanish at the conformal infinity but diverge at the ring singularity. To compute the zeroes of the entropy density function we only need to find the roots of the numerator in (\ref{enden1}), which is a quadratic function in $ r $.

Substituting $ \alpha=0 $ in the above expression of entropy density, we get the entropy density for the Kerr black hole:
\begin{equation}
s_{kerr}=\dfrac{2k_{s}r}{(r^{2}+a^{2}\cos^{2}\theta)}.
\end{equation}

\begin{figure}
\includegraphics[width=0.40\textwidth]{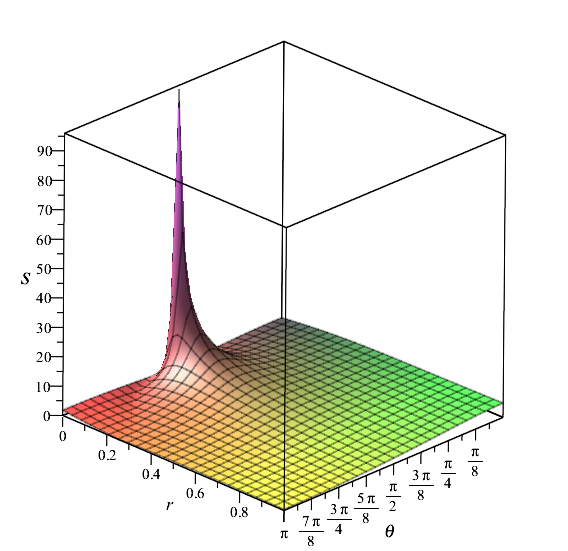}
\caption{Plot showing the variation of the gravitational entropy density for the accelerating rotating BH with respect to the radial coordinate $ r $ and the angular coordinate $ \theta $, where $ \alpha=0.45, \: m=1, \: {\color{blue}{a=0.5,}} \: \textrm{and} \: k_{s}=1 $. This figure clearly indicates that at the ring singularity $\left( r=0, \: \theta=\dfrac{\pi}{2} \right) $ the gravitational entropy density diverges.}\label{plot4b}
\end{figure}

\begin{figure}[ht]
    \centering
    \subfloat[Subfigure 1 list of figures text][]
        {
        \includegraphics[width=0.42\textwidth]{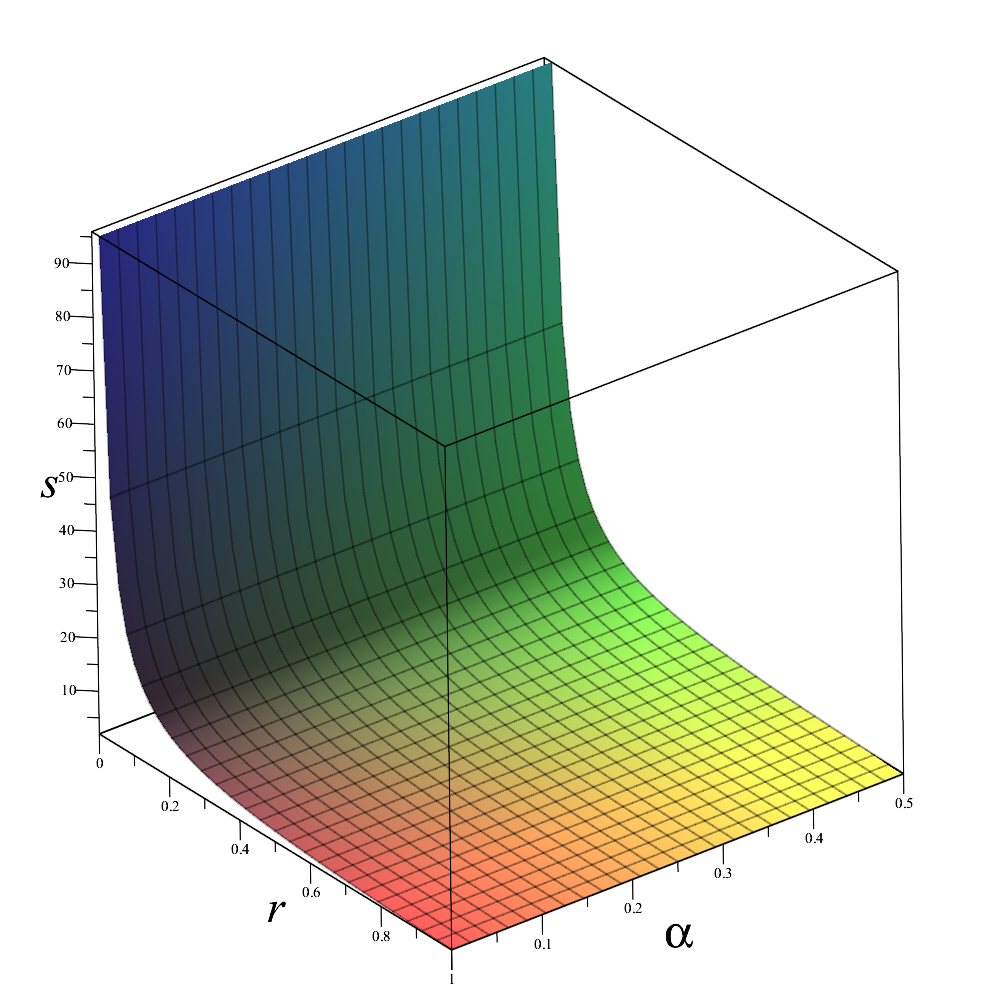}
        \label{fig:subfig7}
        }
        \hspace{1.0cm}
    \subfloat[Subfigure 2 list of figures text][]
        {
        \includegraphics[width=0.37\textwidth]{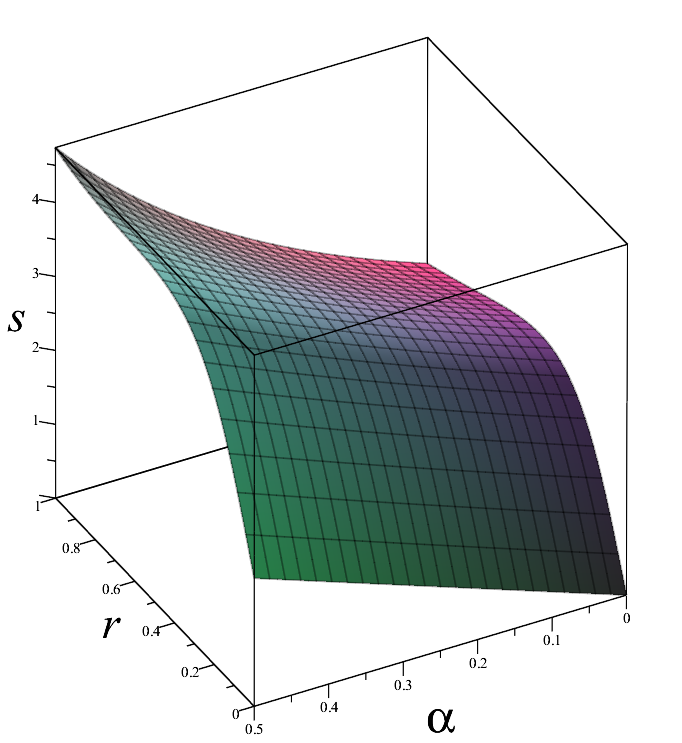}
        \label{fig:subfig8}
        }
    \caption{(a) Plot showing the variation of the gravitational entropy density for the accelerating rotating BH with respect to the radial coordinate $ r $ and the acceleration parameter $ \alpha $, where $ a=0.5, \: m=1, \: k_{s}=1, \: \textrm{and} \: \theta=\dfrac{\pi}{2} $. This figure clearly shows that at the ring singularity $ \left(r=0, \: \theta=\dfrac{\pi}{2}\right) $ the gravitational entropy density diverges. (b) Plot showing the variation of the gravitational entropy density  for the accelerating rotating BH with respect to the radial coordinate $ r $ and the acceleration parameter $ \alpha $, where $ a=0.5, \: m=1, \: k_{s}=1, \: \textrm{and} \: \theta=\dfrac{\pi}{6} $. This figure shows that at $ r=0 $ and $ \theta=\dfrac{\pi}{6} $, the gravitational entropy density is finite.}
    \label{fig4}
\end{figure}

\begin{figure}[ht]
    \centering
    \subfloat[Subfigure 1 list of figures text][]
        {
        \includegraphics[width=0.42\textwidth]{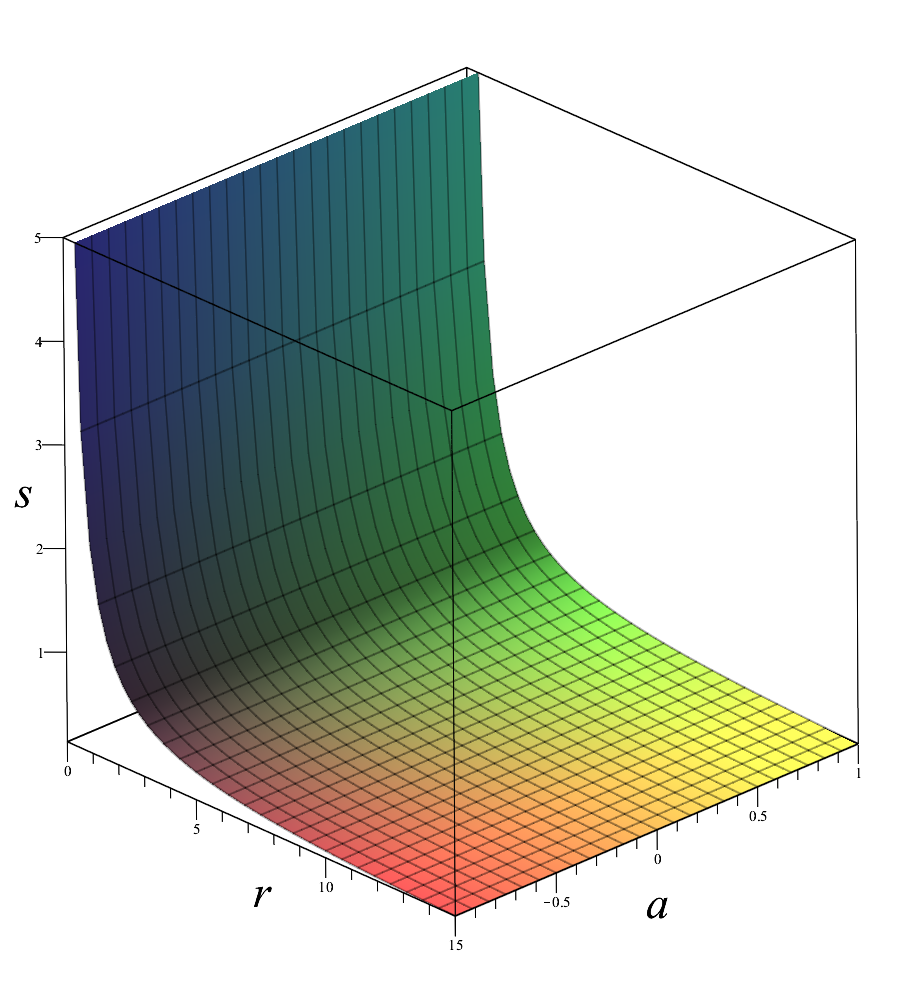}
        \label{fig:subfig9}
        }
    \subfloat[Subfigure 2 list of figures text][]
        {
        \includegraphics[width=0.47\textwidth]{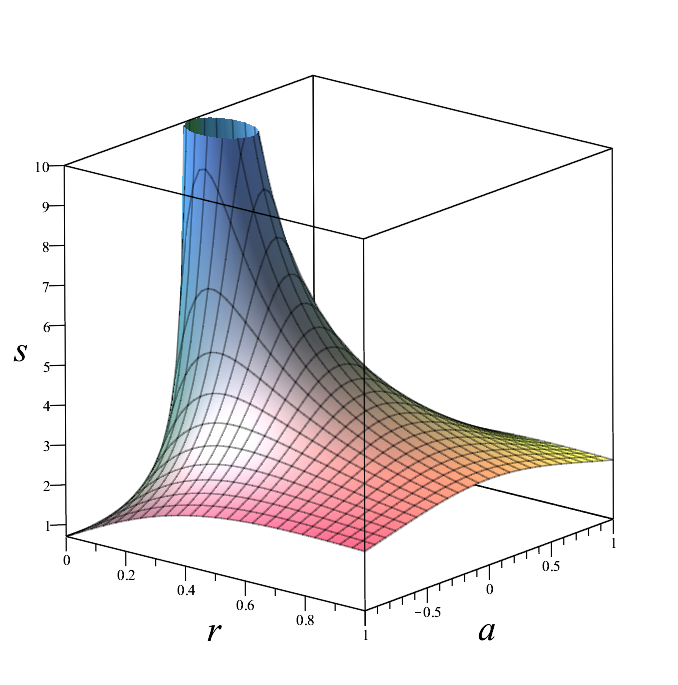}
        \label{fig:subfig10}
        }
    \caption{(a) Plot showing the variation of the gravitational entropy density for the accelerating rotating BH with respect to the radial coordinate $ r $ and the rotation parameter $ a $, where $ \alpha=0.25, \: m=1, \: k_{s}=1, \: \textrm{and} \: \theta=\dfrac{\pi}{2} $. (b) Plot showing the variation of the gravitational entropy density for the accelerating rotating BH with respect to the radial coordinate $ r $ and the rotation parameter $ a $, where $ \alpha=0.25, \: m=1, \: k_{s}=1, \: \textrm{and} \: \theta=\dfrac{\pi}{4} $.}
    \label{fig5}
\end{figure}

\begin{figure}
\includegraphics[width=0.6\textwidth]{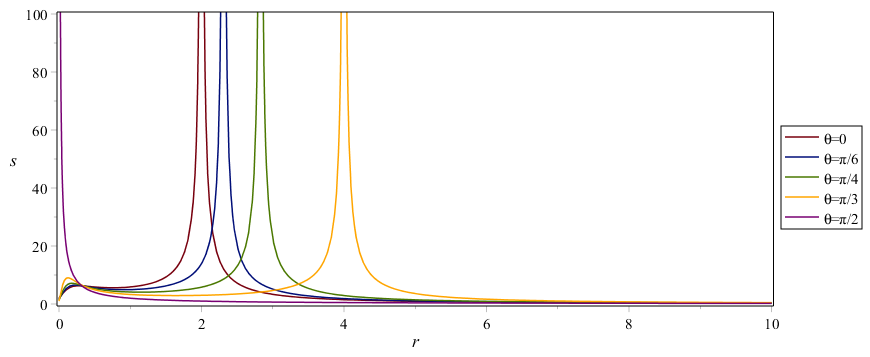}
\caption{Plot showing the variation of the  gravitational entropy density for the accelerating rotating BH with respect to the radial coordinate $ r $ for different values of $ \theta $, where we have taken $m=1 \: a=0.25, \: \alpha=0.5, \textrm{and} \: k_{s}=1 $.}\label{plottn}
\end{figure}

FIG. \ref{plot4b} clearly shows that the measure of entropy density is well behaved everywhere except at the ring singularity. In figures FIG. \ref{fig4}(a) and FIG. \ref{fig4}(b), we have shown that for different values of $\theta$ we can have a diverging or finite entropy density at $r=0$. When $ \theta=\frac{\pi}{2} $, the expression of entropy density in (\ref{enden1}) simply becomes $ \dfrac{2k_{s}}{r} $, which can be seen clearly in FIG. \ref{fig4}(a) and it also diverges at the ring singularity at $ r=0 $ for this case. Whereas in FIG. \ref{fig4}(b), we can see that for $ \theta=\frac{\pi}{6} $ the entropy density is finite at $ r=0 $ for nonzero values of acceleration parameter whereas for  $ \alpha=0 $ the entropy density becomes zero at the central singularity at $ r=0 $. FIG. \ref{fig5}(a) shows us that the entropy density simply behaves like inverse squared in $ r $ when $ \theta=\frac{\pi}{2} $, for different values of $ a $, as it becomes independent of the rotation parameter and the acceleration parameter, which can be easily seen from the expression (\ref{enden1}). In FIG. \ref{fig5}(b), the entropy density diverges for the condition $ r=0, \: \textrm{and} \: a=0 $, because it corresponds to the central singularity of an accelerating non-rotating BH. These behaviors of the entropy density are in conformity with our expectations and so we can say that this definition of entropy density is quite suitable for these kinds of black holes.

In FIG. \ref{plottn}, the nature of the gravitational entropy density for accelerating rotating black hole is studied for different values of $ \theta $, where we have fixed the values of the black hole parameters as the following: $m=1, \: a=0.25, \: \alpha=0.5, \textrm{and} \: k_{s}=1 $. The conformal infinity lies at $ r=\dfrac{1}{\alpha\cos\theta} $. From (\ref{enden1}) it is clear that as the value of $ \theta $ goes from $ 0 $ to $ \dfrac{\pi}{2} $, the conformal infinity shifts towards infinity, giving rise to the $ \sim \dfrac{1}{r^2} $ behavior which diverges only at the ring singularity at $ r=0, \theta=\dfrac{\pi}{2} $. Moreover, we observe that except for $ \theta=\dfrac{\pi}{2} $, the gravitational entropy density remains finite at $ r=0 $ for all other values of $\theta$.

\subsection{Rotating charged accelerating black hole}
For the rotating charged black hole, the Weyl scalar $W$ is
\begin{align}
\left.W\right. &=48\frac { \left( \alpha r\cos \left( \theta \right) -1 \right) ^{6}}
{ \left( {r}^{2}+{a}^{2} \left( \cos \left(\theta \right)  \right) ^{2} \right) ^{6}} \times\nonumber\\
 &(  \left( {e}^{2}r\alpha+am \left( a\alpha+1 \right)  \right) {
a}^{2}\cos^{3}\theta+ \left( 2a{e}
^{2}{r}^{2}\alpha+3{a}^{2}m \left( a\alpha-1 \right) r+{a}^{2}{e}^{2
} \right)   \cos^{2}\theta \nonumber\\
&+ \left( -{e
}^{2}\alpha\,{r}^{3}-3am \left( a\alpha+1 \right) {r}^{2}+2 a{e}^{2
}r \right) \cos \left( \theta \right) -{r}^{2} \left( m \left( a\alpha
-1 \right) r+{e}^{2} \right)  ) \nonumber\\
&  (  \left( {e}^{2}r\alpha+a
m \left( a\alpha-1 \right)  \right) {a}^{2} \left( \cos \left( \theta
 \right)  \right) ^{3}+ \left( -2a{e}^{2}{r}^{2}\alpha-3{a}^{2}m
 \left( a\alpha+1 \right) r+{a}^{2}{e}^{2} \right)  \left( \cos
 \left( \theta \right)  \right) ^{2} \nonumber\\
 &+ \left( -{e}^{2}\alpha{r}^{3}-3
am \left( a\alpha-1 \right) {r}^{2}-2a{e}^{2}r \right) \cos
 \left( \theta \right) +{r}^{2} \left( m \left( a\alpha+1 \right) r-{e
}^{2} \right)  ),
\end{align}
and the Kretschmann scalar $K$ is
\begin{align}
\left.K\right. &=  48\frac {\left( \alpha r \cos \left( \theta \right)-1 \right)^{6}}{\left({r}^{2}+{a}^{2}\left(\cos\left(\theta \right)\right)^{2}\right)^{6}}({a}^{4}\left({a}^{4}{\alpha}^{2}{m}^{2}+2{a}^{2}{\alpha}^{2}{e}^{2}mr+7/6{\alpha}^{2}{e}^{4}{r}^{2}
-{a}^{2}{m}^{2} \right)  \left( \cos \left( \theta \right)  \right) ^{6}  \nonumber\\
&+2 \left( {a}^{2}m+5/6{e}^{2}r \right)  \left( {e}^{2}-6mr\right) {a}^{4}\alpha \left(\cos\left(\theta \right)\right)^{5} \nonumber\\
& +\left(15{a}^{4}{m}^{2}{r}^{2}-20{a}^{4}{\alpha}^{2}{e}^{2}m{r}^{3}-15{a}^{6}{\alpha}^{2}{m}^{2}{r}^{2}-{\frac {17{a}^{2}{\alpha}^{2}{e}^{4}{r}^{4}}{3}}-10{a}^{4}{e}^{2}mr+7/6{a}^{4}{e}^{4}\right)  \left( \cos \left( \theta \right)  \right) ^{4} \nonumber\\
&-20 \left(-{e}^{2}m{r}^{2}+ \left( -2{a}^{2}{m}^{2}+{\frac {19{e}^{4}}{30}}\right) r+{a}^{2}{e}^{2}m \right) {r}^{2}{a}^{2}\alpha \left( \cos\left( \theta \right)  \right) ^{3}+ \nonumber\\
 & \left( 7/6{\alpha}^{2}{e}^{4}{r}^{6}-{\frac {17{a}^{2}{e}^{4}{r}^{2}}{3}}-15{a}^{2}{m}^{2}{r}^{4
}+10{a}^{2}{\alpha}^{2}{e}^{2}m{r}^{5}+15{a}^{4}{\alpha}^{2}{m}^{2}{r}^{4}+20{a}^{2}{e}^{2}m{r}^{3} \right)  \left( \cos \left( \theta\right)  \right) ^{2} \nonumber\\
 &+ 10 \left( {e}^{2}-6/5mr \right) {r}^{4}\alpha \left( {a}^{2}m+1/6{e}^{2}r \right) \cos \left( \theta
 \right) + \left( -{a}^{2}{\alpha}^{2}{m}^{2}+{m}^{2} \right) {r}^{6}-2{e}^{2}m{r}^{5}+7/6{e}^{4}{r}^{4} ).
\end{align}

From the above scalars we can calculate the ratio $ P=\sqrt{\dfrac{W}{K}} $, defined in \cite{entropy1}, which serves as the measure of gravitational entropy, $S_{grav}$ of black holes. The four-dimensional determinant of the metric is
\begin{equation}
g=-{\frac { \left( \sin \left( \theta \right)\right) ^{2} \left( {r}^{2}+{a}^{2} \left( \cos \left( \theta
 \right)  \right) ^{2} \right) ^{2}}{ \left( \alpha\,r\cos \left( \theta \right) -1 \right) ^{8}}}.
\end{equation}
Here again the axisymmetric metric denies us the calculation of the spatial metric due to the nonzero metric component $ g_{t\phi} $.  Therefore as in the previous calculation for rotating black holes, the entropy density is calculated by using the metric determinant $ g $ in the covariant derivative. We thus have
\begin{equation}\label{enden2}
s=k_{s}|\mathbf{\nabla}.\mathbf{\Psi}|=\dfrac{k_{s}}{\sqrt{-g}}\left(\dfrac{\partial}{\partial r}\sqrt{-g}P\right).
\end{equation}
Here we have intentionally avoided writing the exact expression of entropy density as it is lengthy and too much complicated, but we can easily check the validity of the result. We have checked that if we substitute $ e=0 $ in these calculations, then we get back the result for the accelerating rotating black hole.

\begin{figure}[ht]
    \centering
    \subfloat[Subfigure 1 list of figures text][]
        {
        \includegraphics[width=0.38\textwidth]{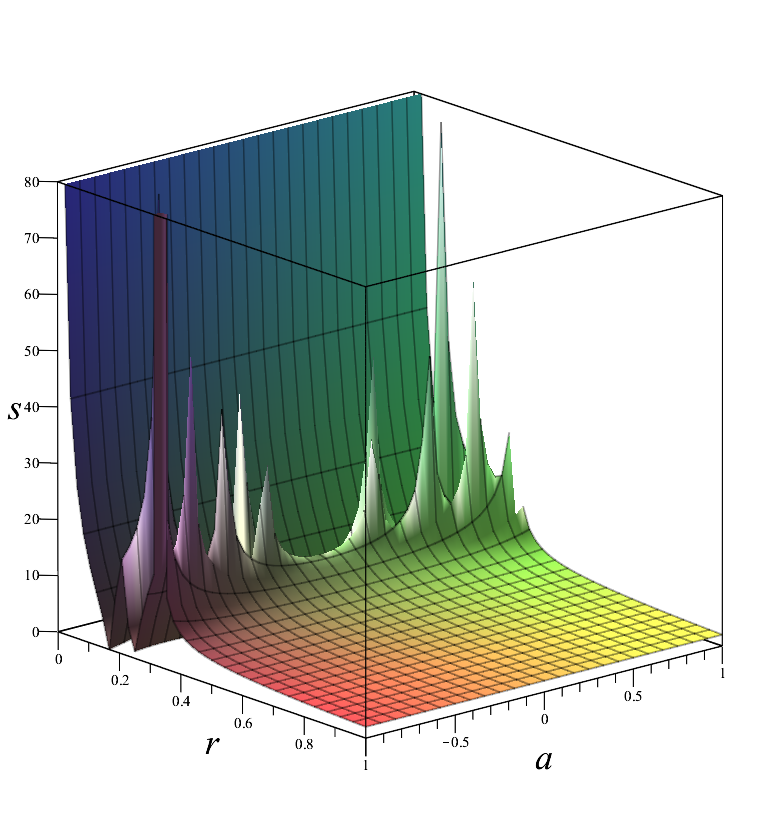}
        \label{fig:subfig11}
        }
        \hspace{1.0cm}
    \subfloat[Subfigure 2 list of figures text][]
        {
        \includegraphics[width=0.4\textwidth]{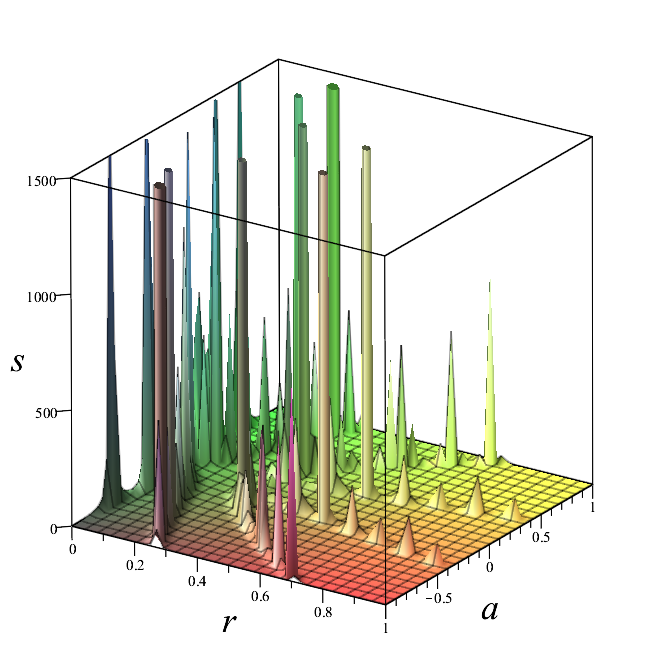}
        \label{fig:subfig12}
        }
    \caption{(a) Plot showing the variation of the gravitational entropy density for the accelerating rotating charged BH with respect to the radial coordinate $ r $ and the rotation parameter $ a $, where $ e=0.6, \: \alpha=0.45, \: m=1, \: k_{s}=1, \: \textrm{and} \: \theta=\dfrac{\pi}{2} $. (b) Plot showing the variation of the gravitational entropy density for the accelerating rotating charged BH with respect to the radial coordinate $ r $ and the rotation parameter $ a $, for $e=0.6, \: \alpha=0.45, \: m=1, \: k_{s}=1, \: \textrm{and} \: \theta=\dfrac{\pi}{4} $.}
    \label{fig6}
\end{figure}
In FIG. \ref{fig6} we find that the gravitational entropy density is not smooth, but contains several singularities. The above analysis clearly shows that the measure of the gravitational entropy used above is not adequate to explain the case of the accelerating rotating charged black holes. Therefore we have to use the measure proposed in \cite{entropy2} for the expression of $ P $, which is
\begin{equation}\label{mod_P}
P=C_{abcd}C^{abcd}.
\end{equation}
Using the definition (\ref{mod_P}) of $ P $, we have calculated the gravitational entropy density, which is given in equation (\ref{new3}):
\begin{align}\label{new3}
\left.s\right.&=\dfrac{k_{s}}{(a^2\cos^2(\theta)+r^2)^7}\bigg(96\Big(a^6\alpha(a^4 \alpha^2 m^2+3a^2\alpha^2 e^2 mr+2\alpha^2 e^4 r^2 -a^2m^2)\cos^9(\theta)\nonumber\\
&+(-20e^2mr^2+(-18a^2m^2+2e^4)r+a^2e^2m)\alpha^2a^6\cos^8(\theta)- \nonumber\\
&34\alpha a^4(21/34e^4r^4\alpha^2+57/34a^2e^2mr^3\alpha^2+(a^4\alpha^2m^2-a^2m^2)r^2+5/34a^2e^2mr-3/17a^4m^2)\cos^7(\theta)+\nonumber\\
&(90a^4e^2mr^4\alpha^2+(142a^6\alpha^2m^2-21a^4\alpha^2e^4)r^3-9a^6e^2m\alpha^2r^2+(20a^8\alpha^2 m^2-20a^6m^2)r+5a^6e^2m)\cos^6(\theta)\nonumber\\
&+30\alpha a^2(8/15e^4r^5\alpha^2+17/6a^2e^2 m r^4\alpha^2+(3a^4\alpha^2 m^2-3a^2m^2)r^3+1/6a^2e^2mr^2\nonumber\\
&+(-19/5a^4m^2+11/30e^4a^2)r+a^4e^2m)r\cos^5(\theta)+(-48a^2e^2m\alpha^2r^6+(-142a^4\alpha^2m^2+16a^2\alpha^2e^4)r^5 \nonumber\\
&-5a^4e^2mr^4\alpha^2+(-90a^6\alpha^2m^2+90a^4m^2)r^3-75a^4e^2mr^2+11a^4e^4r)\cos^4(\theta)-100(1/100e^4r^5\alpha^2+\nonumber\\
&3/20a^2e^2mr^4\alpha^2+(17/50a^4\alpha^2m^2-17/50a^2m^2)r^3-9/100a^2e^2mr^2+(-17/10a^4m^2+13/50e^4a^2)r+ \nonumber\\
&a^4e^2m)\alpha r^3\cos^3(\theta)+(2e^2m\alpha^2r^8+(18a^2\alpha^2 m^2-\alpha^2e^4)r^7+5a^2e^2m\alpha^2 r^6+ \nonumber\\
&(48a^4\alpha^2m^2-48a^2m^2)r^5+75a^2e^2mr^4-26e^4a^2r^3)\cos^2(\theta)+30((1/30a^2\alpha^2m^2-1/30m^2)r^3-1/30e^2mr^2+\nonumber\\
&(-a^2m^2+1/10e^4)r+a^2e^2m)\alpha r^5\cos(\theta)+(-2a^2\alpha^2m^2+2m^2)r^7-5e^2 mr^6+3e^4r^5\Big)(r\alpha  \cos(\theta)-1)^5\bigg).
\end{align}

\begin{figure}[ht]
    \centering
    \subfloat[Subfigure 1 list of figures text][]
        {
        \includegraphics[width=0.46\textwidth]{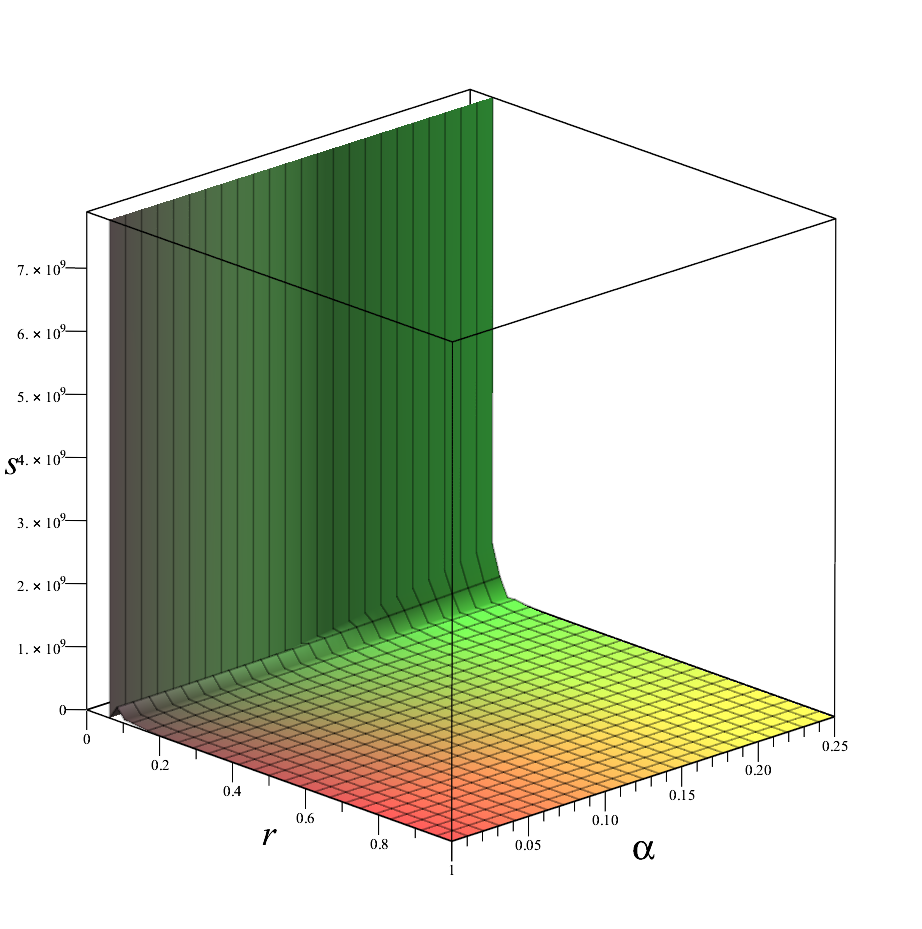}
        \label{fig:subfig13}
        }
    \subfloat[Subfigure 2 list of figures text][]
        {
        \includegraphics[width=0.45\textwidth]{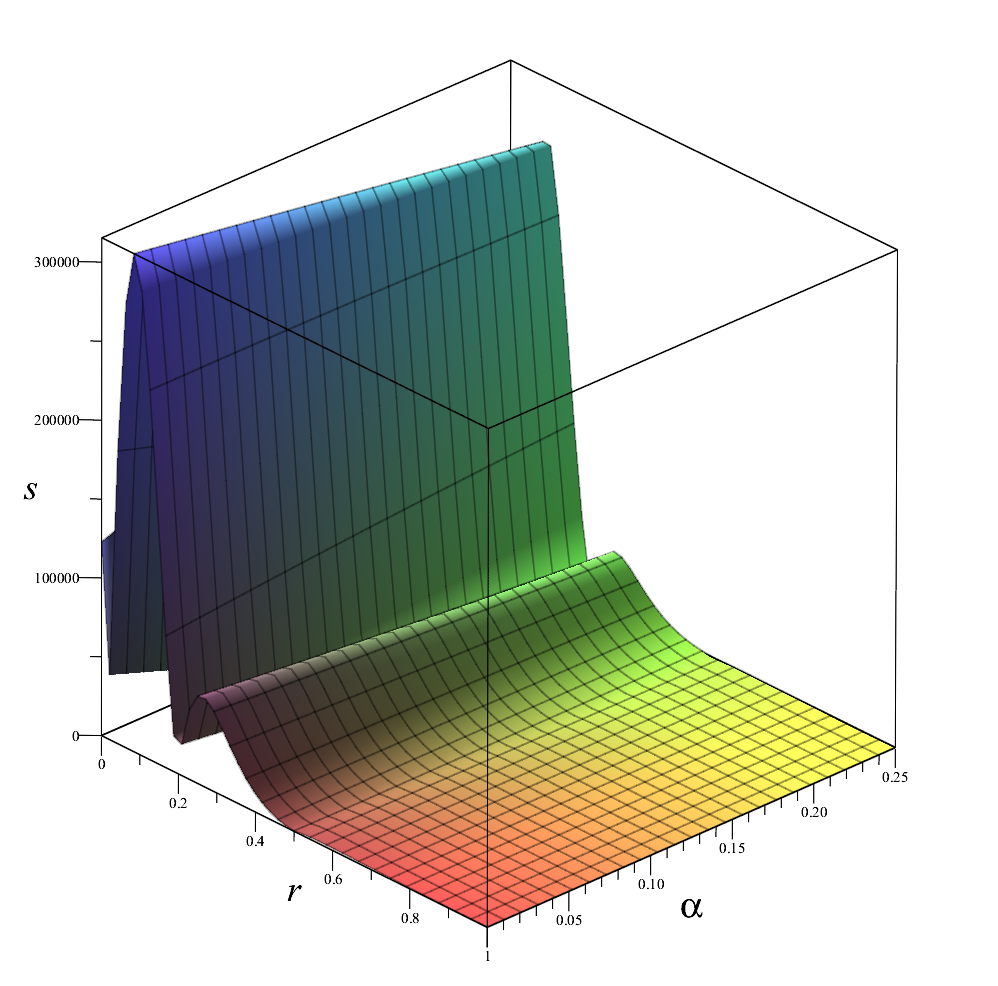}
        \label{fig:subfig14}
        }
    \caption{(a) Plot showing the variation of the gravitational entropy density for the accelerating rotating charged BH with respect to the radial coordinate $ r $ and the acceleration parameter $ \alpha $, for $ e=0.25, \: a=0.5, \: m=1, \: k_{s}=1, \: \textrm{and} \: \theta=\dfrac{\pi}{2} $. (b) Plot showing the variation of the gravitational entropy density for the accelerating rotating charged BH with respect to the radial coordinate $ r $ and the acceleration parameter $ \alpha $, where $e=0.25, \: a=0.5, \: m=1, \: k_{s}=1, \: \textrm{and} \: \theta=\dfrac{\pi}{4} $.}
    \label{fig7}
\end{figure}

In FIG. \ref{fig7} we have shown the variation of the gravitational entropy density with the radial distance and the acceleration parameter using this new definition (\ref{mod_P}) of the scalar $ P $. The entropy density function is now well-behaved and all the singularities vanish, except the ring singularity, on account of the introduction of this new definition. In FIG. \ref{fig7}(a), the entropy density function diverges at $ r=0 $ and $ \theta=\frac{\pi}{2} $, as it encounters the ring singularity, whereas in FIG. \ref{fig7}(b) the entropy density stays finite at $ r=0 $ and $ \theta=\frac{\pi}{4} $. Although the entropy density function (\ref{new3}) vanishes at the conformal infinity $ r=\dfrac{1}{\alpha\cos(\theta)} $, we cannot simply associate the zeroes of the entropy density function with the horizons, because according to this modified definition, the expression (\ref{new3}) does not have such factors, and so we have to solve the function explicitly in order to determine the zeroes of the entropy density.

\section{Discussions}
We now discuss the possibility of having an angular component in the vector field $ \mathbf{\Psi} $ for axisymmetric spacetimes as proposed in \cite{entropy2}. Using this modified definition of $ \mathbf{\Psi} $, and the modified expression (\ref{mod_P}), we now calculate the gravitational entropy density for axisymmetric space-times, using the following expression:
\begin{equation}\label{news}
s=k_{s}|\mathbf{\nabla}.\mathbf{\Psi}|=\dfrac{k_{s}}{\sqrt{-g}}\left|\left(\dfrac{\partial}{\partial r}(\sqrt{-g}P)+\dfrac{\partial}{\partial \theta}(\sqrt{-g}P) \right)\right|.
\end{equation}

\begin{figure}[ht]
    \centering
    \subfloat[Subfigure 1 list of figures text][]
        {
        \includegraphics[width=0.45\textwidth]{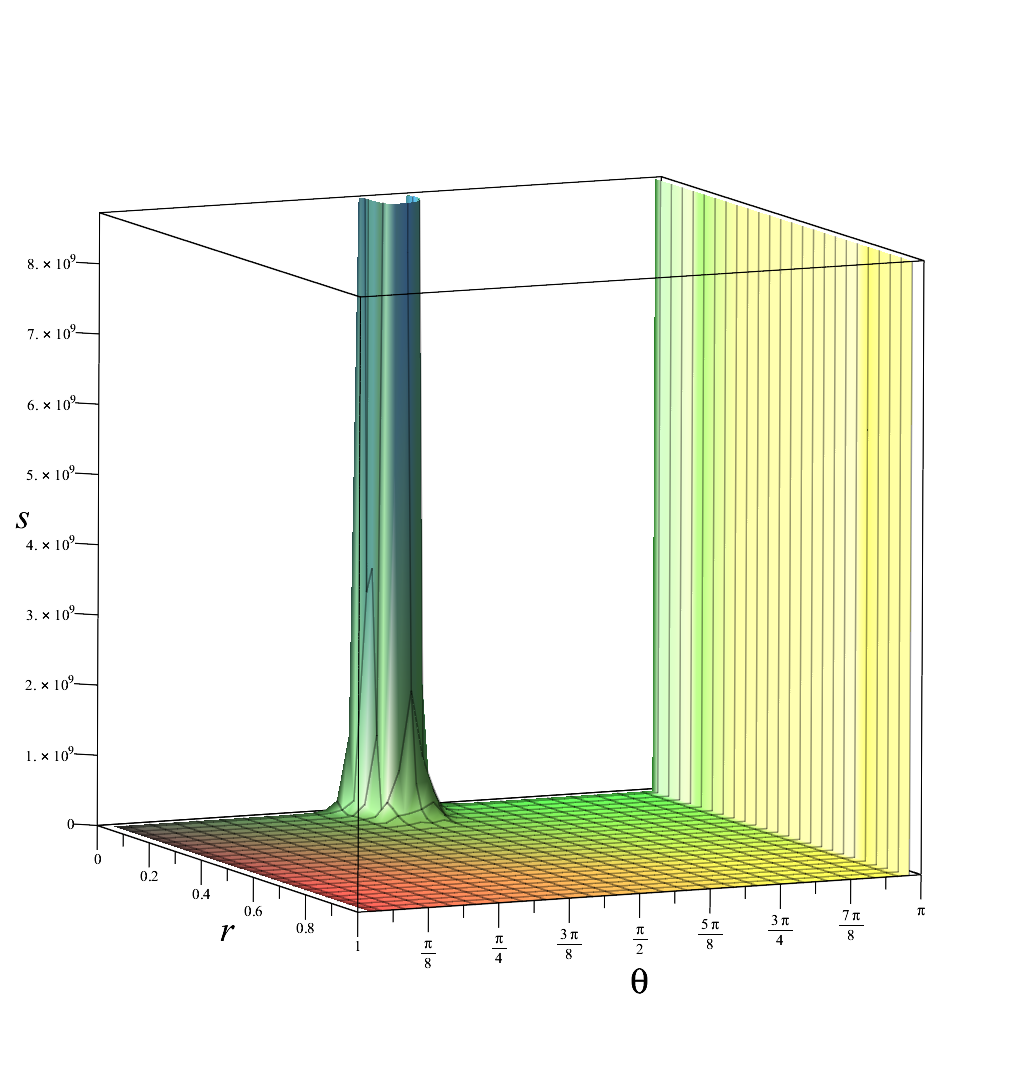}
        \label{fig:subfig15}
        }
    \subfloat[Subfigure 2 list of figures text][]
        {
        \includegraphics[width=0.48\textwidth]{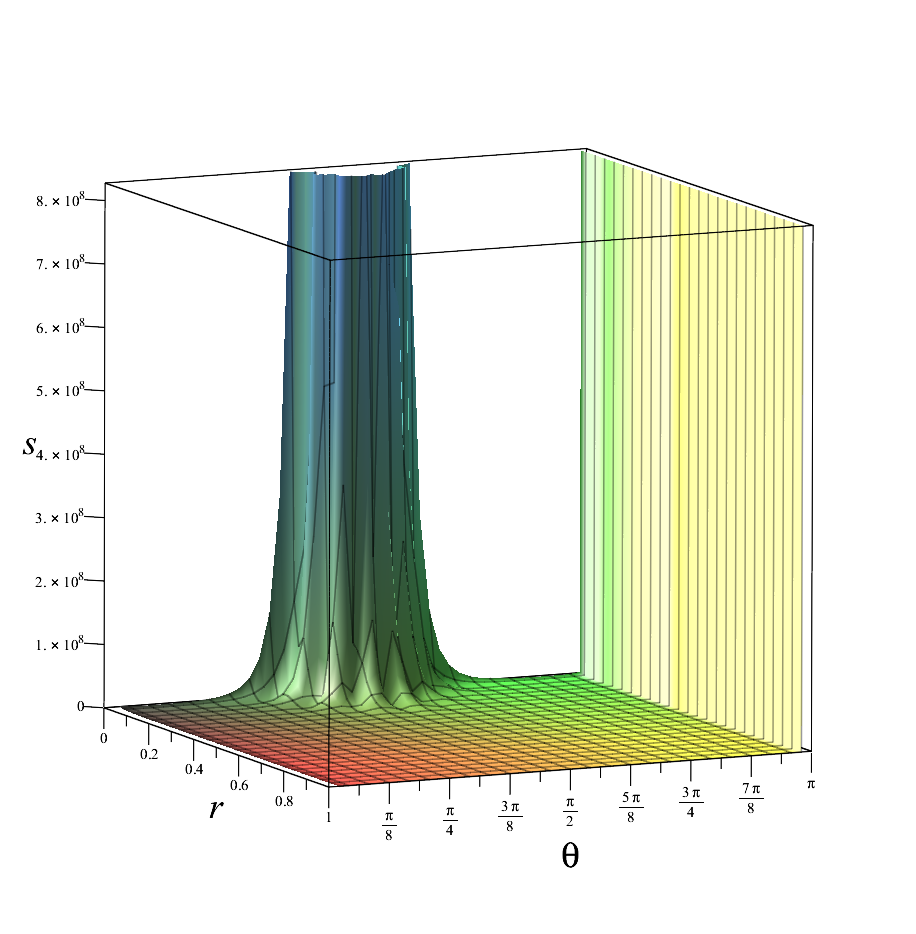}
        \label{fig:subfig16}
        }
    \caption{(a) Plot showing the variation of the gravitational entropy density for the accelerating rotating BH with respect to the radial coordinate $ r $ and the angular coordinate $ \theta $, using the modified expression given in \ref{new1}, where $\alpha=0.25, \: e=0, \: a=0.95, \: m=1, \: \textrm{and} \: k_{s}=1 $. (b) Plot showing the variation of the gravitational entropy density for the accelerating rotating charged BH with respect to the radial coordinate $ r $ and the angular coordinate $ \theta $, using the modified expression given in \ref{new2}, where $\alpha=0.25, \: e=0.2, \: a=0.45, \: m=1, \: \textrm{and} \: k_{s}=1 $.}
    \label{fig8}
\end{figure}

The gravitational entropy density for the uncharged rotating accelerating black hole is given by
\begin{align}\label{new1}
\left.s\right. &= \dfrac{k_{s}}{\sqrt{\dfrac{\sin^2(\theta)(a^2\cos^2(\theta)+r^2)^2}{(r\alpha\cos(\theta)-1)^8}}}\Bigg(\Bigg|\dfrac{48}{\sqrt{\dfrac{\sin^2(\theta)(a^2\cos^2(\theta)+r^2)^2}{(r\alpha\cos(\theta)-1)^8}}(a^2\cos^2(\theta)+r^2)^5(r\alpha \cos(\theta)-1)^3} \nonumber\\
&\bigg(\sin(\theta)\Big(((2a^{10}\alpha^{3} m^{2} r-2a^{8}\alpha m^{2}r)\cos^{8}(\theta)-4(a^{4}m^{2}\alpha^{2}+(9\alpha^{2}r^{2}-1)m^{2}a^{2})a^{6}\cos^{7}(\theta)-10(m^{2}(\frac{34}{5}r^{3}\alpha^{2}  \nonumber\\
&-6r)a^{4}-\frac{34}{5}r^{3}a^{2}m^{2})\alpha a^{4}\cos^{6}(\theta)
+96 a^{4}(a^{4}m^{2}r^{2}\alpha^{2}+(\frac{71}{24}\alpha^{2}r^{4}-r^{2})m^{2}a^{2})\cos^{5}(\theta)+150((\frac{6}{5}r^{3}\alpha^{2}-\frac{34}{15}r) \nonumber\\
&m^{2}a^{4}-\frac{6}{5}r^{3}a^{2}m^{2}) \alpha a^{2}r^{2}\cos^{4}(\theta)-180a^{2}(a^{4}m^{2}r^{2}\alpha^{2}+(\frac{71}{45}\alpha^{2}r^{4}-r^2)m^2a^2)r^2\cos^{3}(\theta)-150\alpha((\frac{34}{75} r^3\alpha^{2} \nonumber\\
&-\frac{38}{25} r)m^2a^4-\frac{34}{75} r^3 a^2 m^2)r^4\cos^{2}(\theta)+40(a^4 m^2 r^2 \alpha^2+(\frac{9}{10} \alpha^2 r^4-r^2)m^2a^2)r^4\cos(\theta)+10\alpha((\frac{1}{5}r^3\alpha^2-\frac{6}{5}r) \nonumber\\
&m a^2-\frac{1}{5}r^3 m)mr^6)\sin^2(\theta)+(2\alpha a^6 (a^4\alpha^2m^2-a^2m^2)\cos^{9}(\theta)-36\cos^8(\theta) a^8 \alpha^2 m^2 r-68 \alpha a^4(m^2(\alpha^2 r^2-\frac{3}{17})a^4 \nonumber\\
& -a^2 m^2 r^2)\cos^7(\theta)+40 a^4(a^4 m^2 r\alpha^2-\frac{9}{20}(-\frac{142}{9} r^3\alpha^2+\frac{20}{9}r)m^2a^2)\cos^6(\theta)+60\alpha a^2((3 r^3 \alpha^2-\frac{19}{5}r)m^2 a^4 \nonumber\\
&-3 r^3 a^2 m^2)r\cos^5(\theta)-180 a^2(a^4 m^2 r^2\alpha^2+(\frac{71}{45} \alpha^2r^4-r^2)m^2a^2)r\cos^{4}(\theta)-200((\frac{17}{50}r^3\alpha^2-\frac{17}{10}r)m^2a^4-  \nonumber\\
&\frac{17}{50}r^3 a^2 m^2)\alpha r^3 \cos^3(\theta)+96 r^3(a^4 m^2 r^2\alpha^2+(\frac{3}{8}\alpha^2 r^4-r^2)m^2a^2)\cos^2(\theta)+60((\frac{1}{30} r^3\alpha^2-r)m^2 a^2-\frac{1}{30}r^3 m^2)\nonumber\\
&\alpha r^5\cos(\theta)-4a^2\alpha^2 m^2 r^7+4 m^2 r^7)\sin(\theta)+(a^2\cos^{2}(\theta)+r^2)(a^2(a^2\alpha m -a m)\cos^3(\theta)+(-3 a^3\alpha m r-3 a^2 m r) \nonumber\\
& \cos^2(\theta)+ (-3 a^2 \alpha m r^2+3 a m r^2)\cos(\theta)+r^2(a\alpha m r+m r))(r\alpha\cos(\theta)-1)(a^2(a^2\alpha m+a m)\cos^3(\theta)+ \nonumber\\
&(3a^3\alpha m r-3a^2mr)\cos^{2}(\theta)+(-3 a^2\alpha m r^2-3 a m r^2)\cos(\theta)-r^3 a m\alpha+r^3 m)\cos(\theta)\Big)\bigg)\Bigg|\Bigg)
\end{align}

In FIG. \ref{fig8}(a), we indicate the variation of the gravitational entropy density with radial distance and the angular coordinate, using the modified expression in (\ref{new1}).

Similarly the gravitational entropy density for the charged rotating accelerating black hole is given by equation (\ref{new2}), which is quoted below. We can always check that if we substitute $ e=0 $ in (\ref{new2}), we get back the expression in (\ref{new1}). In FIG. \ref{fig8}(b), we have shown the corresponding variation of the gravitational entropy density with radial distance and angular coordinate. The expressions of gravitational entropy density in both (\ref{new1}) and (\ref{new2}) diverges at the ring singularity i.e. at $ (r^2 + a^2 \cos^2(\theta))=0 $ and vanishes at the conformal infinity i.e. at $ r=\dfrac{1}{\alpha \cos(\theta)} $.

\begin{align}\label{new2}
\left.s\right. &= \dfrac{k_{s}}{\sqrt{\dfrac{\sin^2(\theta)(a^2\cos^2(\theta)+r^2)^2}{(r\alpha\cos(\theta)-1)^8}}}\Bigg(\Bigg|\dfrac{48}{\sqrt{\dfrac{\sin^2(\theta)(a^2\cos^2(\theta)+r^2)^2}{(r\alpha\cos(\theta)-1)^8}}(a^2\cos^2(\theta)+r^2)^5(r\alpha \cos(\theta)-1)^3} \nonumber\\
&\bigg(\sin(\theta)\Big(((2a^{10} m^2r\alpha^3+(4\alpha^3e^2m r^2-2\alpha m^2 r)a^8+2a^6e^4r^3\alpha^3)\cos^8(\theta)- 4(a^4m^2\alpha^2+((9\alpha^2r^2-1)m^2+\frac{1}{2}e^2mr\alpha^2) \nonumber\\
&a^2 -\frac{1}{2}e^2r^2\alpha^2(e^2-15mr))a^6\cos^7(\theta)-10(m((\frac{34}{5}r^3\alpha^2-6r)m+e^2)a^4+\frac{3}{5}(-\frac{34}{3}m^2r^2+16e^2r(\alpha^2r^2-\frac{5}{48})m+e^4) \nonumber\\
&ra^2 +\frac{16}{5}e^4r^5\alpha^2)\alpha a^4\cos^6(\theta)+96a^4(a^4m^2r^2\alpha^2+((\frac{71}{24}\alpha^2r^4-r^2)m^2+\frac{3}{16}e^2(\alpha^2r^2+\frac{10}{3})rm-\frac{1}{16}e^4)a^2-\frac{1}{3}e^2\alpha^2  \nonumber\\
&(e^2-\frac{85}{16}mr)r^4)\cos^5(\theta)+150(((\frac{6}{5}r^3\alpha^2-\frac{34}{15}r)m+e^2)ma^4+\frac{26}{75}(-\frac{45}{13}m^2r^2+\frac{45}{13}(\alpha^2r^2-\frac{1}{18})e^2rm+e^4)ra^2   \nonumber\\
&+\frac{7}{25}e^4r^5\alpha^2)\alpha a^2r^2\cos^4(\theta)-180a^2(a^4m^2r^2\alpha^2+((\frac{71}{45}\alpha^2r^4-r^2)m^2-\frac{1}{18}e^2r(\alpha^2r^2-20)m-\frac{13}{45}e^4)a^2-\frac{7}{30}e^2\alpha^2r^4  \nonumber\\
&(e^2-\frac{19}{7}mr))r^2\cos^3(\theta)-150\alpha(((\frac{34}{75}r^3\alpha^2-\frac{38}{25}r)m+e^2)ma^4+\frac{11}{75}r(-\frac{34}{11}m^2r^2+\frac{20}{11}e^2(\alpha^2r^2+\frac{9}{20})rm   \nonumber\\
&+e^4)a^2+\frac{2}{75}e^4r^5\alpha^2)r^4\cos^2(\theta)+40(a^4m^2r^2\alpha^2+((\frac{9}{10}\alpha^2r^4-r^2)m^2-\frac{1}{4}e^2r(\alpha^2r^2-6)m-\frac{11}{20}e^4)a^2-\frac{1}{10}(e^2   \nonumber\\
&-\frac{3}{2}mr)e^2\alpha^2r^4)r^4\cos(\theta)+10\alpha(((\frac{1}{5}r^3\alpha^2-\frac{6}{5}r)m+e^2)a^2+\frac{1}{5}r^2(e^2-mr))mr^6)\sin^2(\theta)+(2\alpha a^6(a^4m^2\alpha^2+   \nonumber\\
&(3\alpha^2e^2mr-m^2)a^2+2\alpha^2e^4r^2)\cos^9(\theta)+2\alpha^2a^6((e^2m-18m^2r)a^2+2re^4-20e^2mr^2)\cos^8(\theta)-68\alpha a^4(m^2(\alpha^2r^2  \nonumber\\
&-\frac{3}{17})a^4+(-m^2r^2 +\frac{57}{34}e^2r(\alpha^2r^2+\frac{5}{57})m)a^2+\frac{21}{34}e^4r^4\alpha^2)\cos^7(\theta)+40a^4(a^4m^2r\alpha^2-\frac{9}{20}((-\frac{142}{9}r^3\alpha^2+\frac{20}{9}r)m  \nonumber\\
&+e^2(\alpha^2r^2-\frac{5}{9}))ma^2-\frac{21}{20}e^2\alpha^2(e^2-\frac{30}{7}mr)r^3)\cos^6(\theta)+60\alpha a^2(((3r^3\alpha^2-\frac{19}{5}r)m+e^2)ma^4+(-3r^3m^2+\frac{17}{6}e^2   \nonumber\\
&(\alpha^2r^2+\frac{1}{17})r^2m+\frac{11}{30}re^4)a^2+\frac{8}{15}e^4r^5\alpha^2)r\cos^5(\theta)-180a^2(a^4m^2r^2\alpha^2+((\frac{71}{45}\alpha^2r^4-r^2)m^2+\frac{1}{18}e^2r(\alpha^2r^2+15)m   \nonumber\\
&-\frac{11}{90}e^4)a^2 -\frac{8}{45}e^2r^4\alpha^2(e^2-3mr))r\cos^4(\theta)-200(((\frac{17}{50}r^3\alpha^2-\frac{17}{10}r)m+e^2)ma^4+\frac{13}{50}(-\frac{17}{13}m^2r^2+\frac{15}{26}e^2(\alpha^2r^2  \nonumber\\
&-\frac{3}{5}) rm+e^4)ra^2+\frac{1}{100}e^4r^5\alpha^2)\alpha r^3\cos^3(\theta)+96r^3(a^4m^2r^2\alpha^2+((\frac{3}{8}\alpha^2r^4-r^2)m^2+\frac{5}{48}e^2r(\alpha^2r^2+15)m   \nonumber\\
&-\frac{13}{24}e^4)a^2 -\frac{1}{48}e^2r^4\alpha^2(e^2-2mr))\cos^2(\theta)+60(((\frac{1}{30}r^3\alpha^2-r)m+e^2)ma^2+\frac{1}{10}r(e^4-\frac{1}{3}e^2mr-\frac{1}{3}m^2r^2)) \times   \nonumber\\
&\alpha r^5\cos(\theta) -4a^2\alpha^2m^2r^7+6e^4r^5-10e^2mr^6+4m^2r^7)\sin(\theta)+(a^2\cos^2(\theta)+r^2)(a^2(a^2\alpha m+\alpha e^2r-am)\cos^3(\theta)+     \nonumber\\
&(-3a^3mr\alpha+(e^2-3mr)a^2-2ae^2\alpha r^2)\cos^2(\theta)+(-3a^2mr^2\alpha+(-2e^2r+3mr^2)a-e^2\alpha r^3)\cos(\theta)+r^2(a\alpha mr-e^2   \nonumber\\
&+mr))(r\alpha \cos(\theta)-1)(a^2(a^2\alpha m+\alpha e^2 r+am)\cos^3(\theta)+(3a^3mr\alpha+(e^2-3mr)a^2+2ae^2\alpha r^2)\cos^2(\theta) \nonumber\\
&+(-3a^2mr^2\alpha+(2e^2r-3mr^2)a-e^2\alpha r^3)\cos(\theta)-r^3am\alpha-e^2r^2+r^3m)\cos(\theta)\Big)\bigg)\Bigg|\Bigg)
\end{align}

Therefore from FIG. \ref{fig8}, we find that the entropy density measure diverges not only at the ring singularity but also at $ \theta=\pi $, which renders this measure inappropriate for determining the gravitational entropy in these cases. There is another singularity at $ \theta=0 $, though not visible in FIG. \ref{fig8}, but can be inferred from the mathematical analysis. This is in agreement with the observations in \cite{entropy2} for non-accelerating axisymmetric black holes. This is a disturbing feature of this method of analysis. For a possible resolution of this problem we want to point out that in the case of rotating black holes, the existence of stationary observer is not well defined because of the effect of frame dragging. Nevertheless, we have worked with the chosen definition of gravitational entropy density to get an overall idea of the way things work out. For such cases of axisymmetric space-times, it is not possible to determine the spatial metric $h_{ij} $ because of the presence of the metric coefficient $g_{t\phi}$ in the metric (\ref{ht}) and in metric (\ref{htn}). This is because the object is rotating and the spatial position of each event in the space-time depends on time. Therefore the covariant divergence is calculated from the determinant of the full metric and is given in equations (\ref{enden1}) and (\ref{enden2}).

%---------------------------------------------------------
\section{Conclusions}

In this paper we have adopted a phenomenological approach of determining the gravitational entropy of accelerating black holes as done in \cite{entropy1} and \cite{entropy2}. We find that the gravitational entropy proposal \cite{entropy1} for the accelerating black holes and charged accelerating black holes works pretty well, except for the rotating charged metric where we faced difficulties in this regard. We then considered the alternative definition of $ P $ given in \cite{entropy2} to compute the entropy density and showed that the gravitational entropy is well defined in this case. In the end we considered the vector $ \mathbf{\Psi} $ to have additional angular components for axisymmetric spacetimes, as proposed in \cite{entropy2}, to compute the entropy density for accelerating rotating and accelerating charged rotating black holes. From our calculations and the corresponding plots, we can conclude that for the rotating black holes the entropy density will be well-defined if we change our definition of the vector field $ \mathbf{\Psi} $, be it in the magnitude ($ P $) of it, or in the vector directions (having additional angular components).

\section*{Acknowledgments}
The authors are thankful to the anonymous reviewers for their valuable comments and suggestions. SC is grateful to CSIR, Government of India for providing junior research fellowship. SG gratefully acknowledges IUCAA, India for an associateship and CSIR, Government of India for approving the major research project No. 03(1446)/18/EMR-II.

\end{document}